\documentclass[twocolumn,prl,showpacs]{revtex4}

\usepackage{graphicx}
\usepackage{rotating}
\usepackage{amsmath}
\usepackage{amsfonts}
\usepackage{amssymb}
\usepackage{enumerate}
\usepackage{longtable}
\usepackage{dcolumn}% Align table columns on decimal point
\usepackage{bm}

\begin{document}

\newcommand{\be}{\begin{equation}}
\newcommand{\ee}{\end{equation}}
\newcommand{\bn}{\begin{eqnarray}}
\newcommand{\en}{\end{eqnarray}}

\title{Re-examining the Verwey transition in 
${\rm Fe_{3}O_{4}}$}  

\author{L. Craco$^1$}
\author{M. S. Laad$^2$}
\author{E. M\"uller-Hartmann$^1$}
\affiliation{$^1$ Institut f\"ur Theoretische Physik, Universit\"at zu 
K\"oln, Z\"ulpicher Stra{\ss}e 77, 50937 K\"oln, Germany}
\affiliation{$^2$Department of Physics, Loughborough University, LE11 3TU, UK}

\date{\rm\today}

\begin{abstract}
Motivated by recent structural data questioning the adequacy of the charge 
order (CO)/disorder picture for the Verwey transition (at $T=T_{V}$) in 
magnetite, we re-investigate this issue within a new theoretical picture.  
Using the state-of-the-art LDA+DMFT method, we show that the non-trivial 
interplay between the $B$-site octahedral distortions and strong, 
multi-orbital electronic correlations in the half-metallic state is a 
necessary ingredient for a proper quantitative understanding of the 
physical responses across $T_{V}$.  While weak CO is found to have 
very small effects on the low-$T$ spectral function, the low-$T$ charge 
gap and the resistivity jump across $T_{V}$ are quantitatively reproduced 
only upon inclusion of CO in LSDA+DMFT scheme.  Our results strongly 
suggest that the Verwey transition is dominantly driven by multi-orbital 
electronic correlations with associated JT distortions on the 
$B$-sublattice, and constitutes a non-trivial advance in attempts to 
understand the physics of ${\rm Fe_3O_4}$.

\end{abstract}


\maketitle

Magnetite is a ferrimagnetic spinel oxide with a high Curie temperature, 
$T_{c}=858$~K. Hence, it is viewed as an ideal candidate for room 
temperature spintronic applications. Being the oldest magnetic material 
known to humanity, it has nonetheless defied a consistent physical 
understanding to date. It is believed that the Verwey transition at 
$T_{V}=123$~K~\cite{[1]} is associated with an order-disorder transition 
from a charge ordered (CO) state of $Fe$ ions on the $B$ sublattice
(with $Fe^{2.5+\delta}Fe^{2.5-\delta}$ for $Fe_{B}$ ions) to a 
disordered state ($Fe^{2.5}$ at all $Fe_{B}$ sites)
 at higher $T$. One would expect insulating behavior below 
$T_{V}$, as indeed observed. However, a consistent picture of the electronic 
state and transport has proved to be very controversial.  Photoemission 
(PES) studies lead to conflicting results: Chainani {\it et al.}~\cite{[3]} 
found a finite density of states (DOS) at the Fermi level, in agreement 
with the predicted metallic behavior in the high-$T$ cubic phase. This 
is apparently corroborated by the optical conductivity studies~\cite{[4]}, 
which reveal clear gap opening below $T_{V}$, and significant spectral 
weight transfer at higher energies at higher $T$.  The ``Drude'' 
contribution is estimated to be small, with most of the spectral weight 
spread out over the incoherent part. However, Park 
{\it et al.}~\cite{[5]} use PES and IPES results to conclude that there 
is no spectral weight at $E_{F}$ above $T_{V}$. The gap does not collapse 
across $T_{V}$, but only shrinks by about $50~meV$.  This would be consistent 
with a conductivity jump by a factor of $100$ across $T_{V}$ only if 
semiconducting behavior is assumed above $T_{V}$ as well.  Qualitatively 
similar conclusions have been derived very recently by Schrupp 
{\it et al.}~\cite{[6]}, who, moreover, use soft X-ray photoemission to 
determine the bulk electronic structure (PES lineshape).  Thus, the Verwey 
transition is a semiconductor to semiconductor transition, rather than 
an insulator-metal transition, as thought earlier.  Other observations 
like the small carrier mobility, the unusual $T$-dependence of the 
conductivity, and the mid-infrared peak in optics~\cite{[4]} suggest 
electronic localisation, but now in both phases.   

The existence and nature of the CO state in ${\rm Fe_{3}O_{4}}$ is equally 
controversial.  New high resolution powder diffraction data~\cite{[7]} find 
very small charge disproportionation, $-0.1<\delta<0.1$, while resonant 
X-ray scattering studies even concluded the absence of CO below 
$T_{V}$~\cite{[8]}.  Thus, contrary to traditional expectation, the 
electrostatic minimisation required for the CO cannot be the dominant mechanism
driving the Verwey transition.  On the other hand, the possibility of 
orbital order (OO), perhaps of a more complex type, has not hitherto been 
investigated.  Recall that $Fe^{2+}$ is a Jahn-Teller active ion, and, 
in fact, that the low-$T$ distortions of the $O$ octahedra around $B$ site
ions can be viewed as JT distortions~\cite{[7]}.  Indeed, a recent LDA+U 
calculation finds that the local occupation of $t_{2g}$ orbitals on
$B$ sites follows the octahedral distortion expected from a JT effect.  
They also find a charge disproportionation close to the experimentally
suggested (small) value, but this could probably be washed out by dynamical 
frustration effects beyond LDA+U in the real system. Observation of the 
distorted octahedra around $B$ sites in the refined low-$T$ structure 
suggests that orbital correlations and associated JT effects in a 
fully magnetically polarised situation might be essential ingredients 
for understanding the Verwey transition.  

An important feature is clear from PES, IPES and optical measurements across
$T_{V}$.  Characteristic strong correlation signatures are clearly visible.  
In particular, the detailed forms of the lineshapes in each phase, as well 
as the appreciable transfer of dynamical spectral weight over large energy 
scales across $T_{V}$ constitute clinching evidence for a phase transition 
between two strongly correlated phases accompanied by strong spectral 
weight transfer (SWT). A direct comparison with LSDA(+U)~\cite{[14]} results 
clearly shows substantial discrepancy between theory and experiment over 
the whole energy range, pointing to the basic inadequacy of LSDA(+U) to 
describe the actual, correlated electronic structure.  This mandates 
extension of LSDA to include dynamical effects via LSDA+DMFT~\cite{Kot}.

Previous theoretical work has dealt with this problem on two fronts.  
Various bandstructure calculations~\cite{[11],[12]} indicate a 
half-metallic state with a gap in the majority density-of-states (DOS).  
This conflicts, however, with the observed semiconducting resistivity 
up to $T=320~K$, and so recent LDA+U work~\cite{[14]} attempts to cure 
this malady.  Moreover, the role of the $B$-site octahedral distortion
has been clarified in the LDA+U work, which suggests that the Verwey 
transition may be understandable as an orbital order/orbital disorder
transition.  Experimentally, the situation is far from clear, and 
theoretically, one expects staggered (Neel) orbital order to be 
susceptible to strong dynamical frustration (in a way similar to 
what happens to usual antiferromagnetism on inverse spinels).  
However, the possibility of more complex orbital ordering patterns, 
involving more than one tetrahedron, is not ruled out in principle.  
On the second front, Ihle and Lorenz~\cite{[16]} explicitly considered 
a scenario involving a phase transition from a CO insulator below $T_{V}$ 
to a short-range ordered, ``metallic'' state above $T_{V}$. This model 
gives rise to polaronic band motion above $T_{V}$ along with polaronic 
hopping conductivity at higher $T$, in good agreement with observations.  
However, in light of experiments questioning the relevance (or even 
existence) of CO, a new starting point may be necessary. Motivated  by 
new structural data~\cite{[7]} and by the LSDA+U work~\cite{[14]}, an 
order-disorder transition in the $t_{2g}$ orbital sector on the 
$B$-sublattice ($Fe^{2+}-Fe^{3+}$), along with concomitant Jahn-Teller 
distortions, might be an attractive candidate for a consistent 
understanding of this system.  We point out that study of the effects 
of orbital correlations/JT distortions in geometrically frustrated TM 
oxides is an enterprise in its infancy~\cite{[17]}; these studies show 
that the combination of strong correlations in the $d$ shell and 
geometrical frustration can induce qualitatively new, complex behaviors.

Summarising, the microscopic origin of carrier localisation, and of its 
modification across $T_{V}$, remains an open issue. The clarification of 
this issue should go hand-in-hand with a microscopic understanding of the 
changes in electronic structure across the transition. 
 Here, we explore this new proposal in detail.  We marry the 
state-of-the-art LSDA~\cite{[12]} with multi-orbital dynamical mean field 
theory [(MO)DMFT] restricted to the $t_{2g}$ sector (see below). We choose 
LSDA+(MO)DMFT because though LSDA captures structural effects reliably in 
the detailed one-electron dispersion(s), it generically fails to yield the 
correlated, insulating/metallic ground states characterised by appreciable 
dynamical SWT.  

We start with the recent LSDA density-of-states (DOS)~\cite{[12]}. 
This describes a half-metallic ferromagnet, with full 
spin polarisation over an appreciable energy scale around $E_{F}$. The 
total DOS has dominant $Fe(B)$ site contributions, but those coming from 
$Fe(A)$ sites, as well as $O$-2p states cannot be neglected either.  
Strictly speaking the $Fe$ DOS also has contributions coming from $d-p$ 
hybridisation.  In what follows, we include only the $Fe-t_{2g}$ DOS 
as the bandstructure (LDA) input in our calculation.  The doublet $E_{g1}$ 
orbitals lie about $0.019~eV$ above the singlet $A_{1g}$ orbital, giving a 
$t_{2g}$ level splitting, $\Delta_{t_{2g}}=\Delta=0.019~eV$. The $Fe-e_{g}$ 
states, split by the crystal field, lie much higher and can be safely 
neglected. {\it Ab initio} estimates yield the intra-orbital Hubbard 
$U\simeq 3.7-4.0~eV$, and, with a Hund coupling, $J_{H}\simeq 1.0~eV$, the 
inter-orbital Hubbard interaction, $U'\simeq (U-2J_{H})=1.7-2.0~eV$.  
Further, $\Delta$ acts like an external 
field in the orbital sector, sensitively controlling the occupations of 
the $t_{2g}$ orbitals (orbital polarisation) in much the same way as the 
magnetisation of a paramagnet is a function of an external magnetic field.  
In addition, the nearest neighbor Coulomb interaction, $V$, between n.n 
$Fe$ sites is important~\cite{[Seo]} in ${\rm Fe_{3}O_{4}}$: its value 
is not reliably known, but is estimated to be $V\simeq 0.4~eV$.  With 
this, the total many-body Hamiltonian for ${\rm Fe_{3}O_{4}}$ is,
\bn
\nonumber
H &=& \sum_{{\bf k}ab\sigma}\varepsilon_{\bf k}^{ab} 
c_{{\bf k}\sigma}^{\dag a}c_{{\bf k}\sigma}^b + 
U\sum_{ia}n_{i\uparrow}^a n_{i\downarrow}^a 
+U'\sum_{iab}n_{i}^a n_{i}^b \\ \nonumber 
&+& V\sum_{<i,j>ab}n_{i}^a n_{j}^b
-J_{H}\sum_{iab}{\bf S}_{i}^a \cdot {\bf S}_{i}^b 
+ \Delta\sum_{iab}(n_{i}^a -n_{i}^b )\;,
\en
where $a,b$ label the $t_{2g}$ orbitals.

We use (MO)DMFT to solve this three-orbital model. Further, the iterated 
perturbation theory (IPT) is used as an ``impurity solver''.  While not 
exact, IPT does provide very good agreement with QMC results for the 
one-band Hubbard model at high-$T$, and with exact diagonalisation as 
well as very recent dynamical-DMRG results for the one-particle spectral 
functions~\cite{[KR],[KH]}.  The detailed formulation of (MO)IPT for 
orbitally degenerate systems has already been developed and 
used~\cite{[LC]}, so we do not repeat the equations here.  Since 
${\rm Fe_{3}O_{4}}$ is half-metallic, we use the (MO)IPT version developed 
earlier by us for ${\rm CrO_{2}}$~\cite{[LC]} to study the spectral function
in the minority spin sector.  We extend earlier work to include the effect of
$V$ by decoupling it in the Hartree approximation, which is exact as 
$d\rightarrow\infty$.

\begin{figure}[t]
%\begin{center}
\includegraphics[width=\columnwidth]{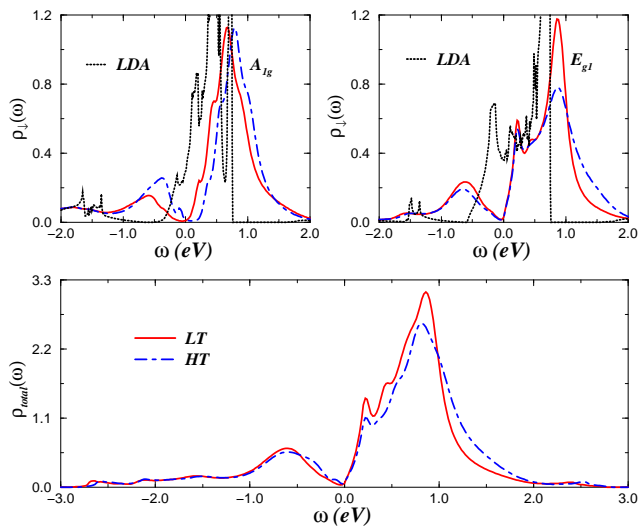}
%\includegraphics[width=3.1in]{f1-p31.eps}
%\end{center}
\caption{(Color online) Orbital-resolved (upper panels) and total 
(lower panel) one electron spectral functions for the two insulating 
phases of ${\rm Fe_3O_4}$.  Notice the changes in the $A_{1g}$ orbital DOS; 
the $E_{g1}$ orbitals shows small changes, demonstrating the nearly 
{\it two-fluid} character of the Verwey transition.}
\label{fig1}
\end{figure}

For ${\rm Fe_{3}O_{4}}$, we start with the high-$T$ cubic structure, with its 
bandstructure computed by LSDA~\cite{[12]}. In addition to generating 
multi-orbital Hartree shifts, which statically renormalise the LDA orbital 
energies by amounts depending upon $U',J_H$ as well as by the occupation of
each $t_{2g}$ orbital, the second order diagrams in DMFT explicitly
account for the dynamical nature of electronic correlations, leading to 
drastic shifts in spectral weights and non-trivial changes in the LDA 
lineshapes.  With $U'=1.7eV, V=0.4eV$, we derive the correlated 
spectral function in the high-$T$ phase: the result is shown in 
Fig.~\ref{fig1}. The (Mott) insulating behavior is readily apparent, as 
is the drastic modification of the LSDA spectrum by multi-orbital 
dynamical correlations.  We estimate the (Mott) gap by extrapolating the 
leading edge of the total DOS to the $\omega$ axis, obtaining 
$\Delta_{g}^{HT}=0.051~eV$, in good agreement with experimental indications.

Motivated by the possible absence of CO in ${\rm Fe_{3}O_{4}}$, and by 
the LSDA+U work showing up the important role of the JT distortion in 
the $B$-sublattice, we adopt the following strategy to study the Verwey 
transition.  We start by noticing that, in a strongly correlated system, 
the correlated values of structural parameters {\it do not} have any 
correspondence with bare LDA values anymore; in fact, these bare LDA 
parameters do not have a clear physical meaning when correlations are 
strong, and are renormalised in widely unanticipated ways by multi-orbital 
interactions. So we adopt a new strategy, where we seek the instability 
of the high-$T$ phase as a function of the $B$-site JT distortion. We 
vary the crystal field on the $B$ sites in small trial steps, starting 
from its LDA+DMFT value (and not the bare LDA value) in the high-$T$ 
cubic phase. For each trial value, we monitor the corresponding 
spectral functions, as well as the renormalised values of orbital 
occupations and the $B$-site distortion from the correlated solution, 
in a way similar to that done for ${\rm V_{2}O_{3}}$ and 
${\rm VO_2}$~\cite{[ML]}. Our procedure is conceptually very different 
from that used in other works~\cite{Held}, where the two correlated phases 
are described by two separate LDA+DMFT calculations, with the corresponding 
LDA DOS for {\it each} phase as inputs into the DMFT calculations.        
          
\begin{figure}[t]
%\begin{center}
\includegraphics[width=\columnwidth]{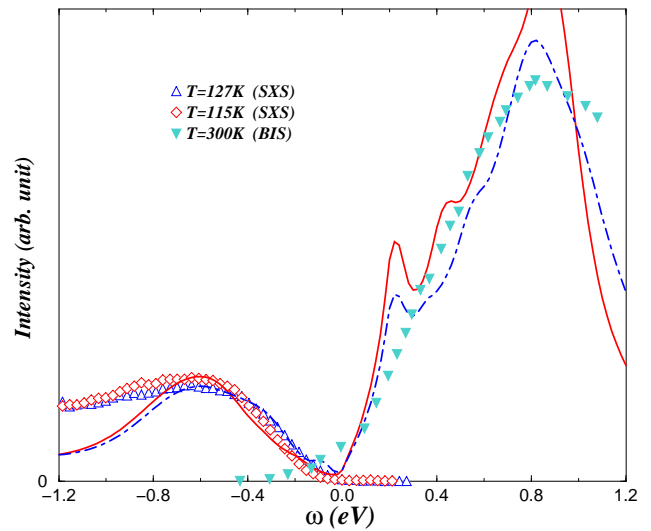}
%\includegraphics[width=3.1in]{f2-p31.eps}
%\end{center}
\caption{(Color online) Comparison of theoretical 
LSDA+DMFT results for the total %one-electron 
spectral function in the high- and low-T
phases of ${\rm Fe_3O_4}$  with experimental results taken from 
Ref.~\protect\onlinecite{[6]} (for PES) and from 
Ref.~\protect\onlinecite{[5]} (for BIS).}
\label{fig2}
\end{figure}

We use this strategy to investigate the Verwey transition in magnetite.  
Beginning from the LSDA+DMFT value of $\Delta=E_{A_{1g}}-E_{E_{g1}}$ 
as the initial input, we selfconsistently search for the critical value 
of $\Delta(=\Delta_{c})$ which destabilises the high-$T$ phase, within 
LSDA+DMFT. We find, as will become clearer below, that the second 
solution of the DMFT equations becomes more stable for $\Delta_c=0.01~eV$.  
Interestingly, (lower panel of Fig.~\ref{fig1}), the second solution also 
corresponds to an insulator, but with a larger gap $O(0.1~eV)$: we point 
out that the low-$T$ phase is indeed found to have a larger gap in 
PES~\cite{[5],[6]}. As for the high-$T$ phase, dynamical multi-orbital 
correlations drastically modify the LSDA(+U) spectra in the low-$T$ phase 
as well.

We now compare our LSDA+DMFT results with recent published 
PES~\cite{[5],[6]} and BIS~\cite{[5]} measurements (BIS was done only 
for the high-$T$ phase). Remarkably, excellent quantitative agreement 
with the occupied as well as unoccupied parts of the spectrum is clearly 
revealed (Fig.~\ref{fig2}) over the whole energy range from $-0.7~eV$ to 
$1.0~eV$.  Concomitantly, comparing the two LSDA+DMFT solutions, we 
find that the high-$T$ phase is also insulating with {\it smaller} gap 
($\Delta_{g}^{HT}=0.051~eV$) than that for the low-$T$ phase 
($\Delta_{g}^{LT}=0.075~eV$). This is again in agreement with the 
PES spectra.  

\begin{figure}[t]
%\begin{center}
\includegraphics[width=\columnwidth]{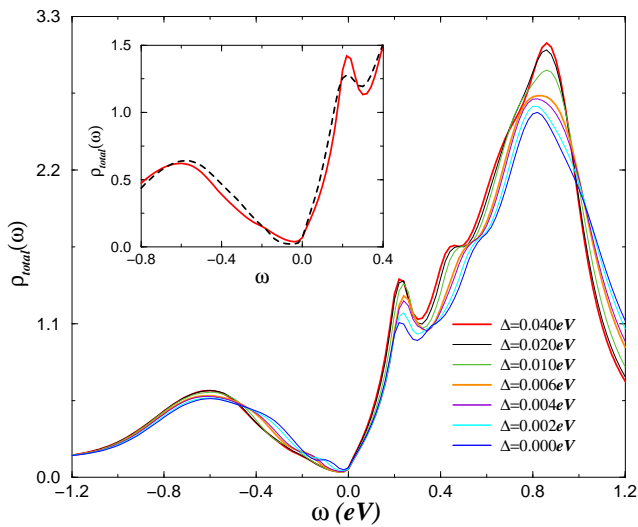}
%\includegraphics[width=3.1in]{f3-p31.eps}
%\end{center}
\caption{(Color online) Effect of the $t_{2g}$ level splitting, $\Delta$ 
on the total one electron spectral function across the Verwey transition 
in ${\rm Fe_3O_4}$. Inset shows the effect of charge order (dashed line) 
in the low-T phase.}
\label{fig3}
\end{figure}

Further light on the character of the Verwey transition is shed by 
studying the evolution of the full many body DOS as a function of $\Delta$. 
In Fig.~\ref{fig3}, we show how the DOS depends on $\Delta$ within 
LSDA+(MO)DMFT. It is indeed remarkable that our results resemble the 
experimental ones~\cite{[6]} very closely if we plausibly suppose that 
$\Delta$ changes with $T$. Up to a critical value $\Delta_{c}$ of the 
crystal field splitting, all curves lie essentially on the one corresponding 
to the high-$T$ phase. For $\Delta > \Delta_{c}$, however, we clearly see 
that all the curves seem to collapse onto the result for the low-$T$ phase.  
Appreciable changes in dynamical spectral weight transfer occur across 
$\Delta_{c}$, in full quantitative accord with observations.  This implies 
that the charge gap increases rapidly as $\Delta$ is raised above 
$\Delta_{c}$, closely tracking the resistivity jump across $T_{V}$.

Up to this point, our calculation does not require the explicit inclusion 
of charge order (CO) within DMFT. This has long been thought to be essential 
in ${\rm Fe_{3}O_{4}}$, but, as argued recently (see above), even the 
existence of CO has proved to be controversial. In any case, if a small 
charge disproportionation would indeed exist, this would result in the 
addition of a minute staggered ``field'' (coming from $Vn_{i}^a n_{j}^b$ 
in $H$), or staggered chemical potential term to the DMFT Hamiltonian.  With 
$V=0.4~eV$ and disproportionation $\eta=0.1$~\cite{[7]}, the staggered 
field term is $H_{s}=V\delta(-1)^{i}n_{i}^a=0.04(-1)^{i}n_{i}^a$; 
we expect that this small 
correction will result in miniscule changes in the electronic structure.
However, using single site CPA to treat dynamical effects of CO in the 
low-$T$ phase, we compute a small correction to the gap (at low-$T$)
of order $0.04\langle n_{i}^b\rangle \simeq 0.012$ (with 
$\langle n_{i}^b \rangle=0.255$ for the $E_{g1}$ orbitals).  From the 
inset of Fig.~\ref{fig3} the low-$T$ 
gap is now estimated to be $\Delta_{g}^{LT}=0.087~eV$.    
The smallness of this term leads us to conclude that the electrostatic 
energy gain associated with the CO would be too small to drive the Verwey 
transition. Moreover, the insulating behavior, as seen in the $dc$ 
resistivity (and fitted previously by an activated hopping term) is also 
naturally accounted for: assuming this form, 
$\rho_{dc}(T) \simeq exp(\Delta_{g}/k_{B}T)$ for both phases  at $T_{V}$
with the appropriate gaps derived above, we find that the resistivity 
jump across $T_{V}$ is $r=e^{4.5} \simeq 90$.  This is very close 
to the factor of slightly less than $100$ as estimated 
experimentally~\cite{[Balberg]}, constituting a further check on our results. 

 In conclusion, in line with very recent experiments, we propose that 
$B$-site JT distortions in a correlated multi-orbital situation play a 
crucial role in understanding the changes in electronic structure across 
the Verwey transition in ${\rm Fe_{3}O_{4}}$.  In the strongly correlated 
situation, small changes in the $B$-site octahedral distortions with $T$ 
give rise to large changes in dynamical spectral weight transfer from low- 
to high energies as $T$ is raised. The excellent quantitative agreement 
with the experimental spectral function, as well as with the gap values 
in {\it both} phases and with the resistivity jump across $T_{V}$ 
constitutes strong support for our new mechanism involving the role of 
multi-orbital correlations and associated structural effects in the 
half-metallic situation. 
 
\acknowledgements
The authors would like to acknowledge A. N. Yaresko for valuable 
discussions and for providing the LSDA DOS.
LC's work was done under the auspices of the Sonderforschungsbereich 
608 of the Deutsche Forschungsgemeinschaft. 
MSL acknowledges financial support from the EPSRC (UK).

\end{document}